%% file: sharuecomm2.tex
\documentstyle[aps,twocolumn,epsfig,amstex,amssymb]{revtex}	

\input{mymath}

\input{mymhd}

\newcommand{\tout}{{\text{out}}}
\newcommand{\tin}{{\text{in}}}
\renewcommand{\Re}{{\text{Re}}}

\begin{document}

\title{Comment on ``The linear instability of magnetic Taylor-Couette flow with Hall effect''}

\author{M. Rheinhardt}
\address{Astrophysikalisches Institut Potsdam,
         An der Sternwarte 16, 14482, Potsdam, Germany, mreinhardt@@aip.de}
\author{U. Geppert}
\address{Max-Planck-Institut f\"ur Extraterrestrische Physik, Gie{\ss}enbachstra{\ss}e, 85748 Garching, Germany, urme@@xray.mpe.mpg.de}

\date{\today}

\maketitle

\begin{abstract}
In the paper we comment on (R\"udiger \& Shalybkov, Phys. Rev. E. 69, 016303 (2004) (RS)), the instability of the Taylor--Couette flow interacting with 
a homogeneous background field subject to Hall effect is studied.
We correct a falsely generalizing interpretation of results presented there which
could be taken to disprove the existence of 
the Hall--drift induced magnetic instability described in Rheinhardt and Geppert, Phys. Rev. Lett. 88,
101103. It is shown that in contrast to what
is suggested by RS, no additional shear flow is necessary
to enable such an instability with a non--potential magnetic background field,
whereas for a curl--free one it is. In the latter case, the instabilities found in RS in situations where neither
a hydrodynamic nor a magneto--rotational instability exists are demonstrated to be 
most likely magnetic instead of magnetohydrodynamic.
Further, some minor inaccuracies are clarified. 
\end{abstract}

\section{}
The main purpose of this Comment on the paper \cite{RS03} (further on referred to as RS) is to prevent a incorrect
conclusion with respect to our work
\cite{RG02,GR02} which could be drawn from an incorrect statement in the discussion
section of RS. There, at the end of the third paragraph, 
the authors conclude from the invariance of their results
with respect to simultaneous sign inversions of shear and Hall term that no instabilities
are possible without shear. Although this conclusion being looked at out of context 
is not comprehensible, it is nevertheless true for the special case of a homogeneous
(more generally: curl--free)
background field $\Bvec_0$, but not in general. As the scheme (40) of RS is valid 
for nonpotential (axisymmetric) fields, too, and the quoted conclusion is drawn completely
on its basis, the reader will be tempted to generalize it.
He or she could then come to the end that the results on a Hall instability {\em without} shear
reported in \cite{RG02,GR02} have to be put in question. (Note, that the term `shear'
is used throughout RS to refer to the macroscopic motion of a fluid.)  
Here, we will show that conclusions on necessary conditions for the instabilities in 
question can reliably be drawn on the basis of energy considerations. They support
the possibility of a Hall instability without shear.

The linearized induction and Navier--Stokes equations
describing the
evolution of small perturbations $\Bvec'$ and $\uvec'$ of the background field
$\Bvec_0$ and the shear flow (here: differential rotation)
$\uvec_0$, respectively, read for a curl--free $\Bvec_0$ 
\begin{equation}
\begin{aligned}
 \parder{\Bvec'}{t} = \phantom{- \beta}&\curl(\uvec_0\times\Bvec'+\uvec'\times\Bvec_0) + \eta\lapl\Bvec'\\ 
 - \beta&\curl(\curl\Bvec'\times\Bvec_0)
\end{aligned}\label{indgl}
\end{equation}
\begin{equation}
\begin{aligned}
 \parder{\uvec'}{t} &+(\uvec'\nabla)\uvec_0 + (\uvec_0\nabla)\uvec' = \\ 
 &- \nabla p'/\rho + \nu\lapl\uvec' + \curl\Bvec'\times\Bvec_0/(\mu_0\rho)  
\end{aligned}\label{navstok}
\end{equation}
where we used the symbols introduced in RS. Standard
arguments yield the following evolution equation for the total energy $E$ of the perturbations:
\begin{equation}
\hspace*{-1mm}\begin{aligned}
&\order{E}{t}=\rezip{2}\order{}{t}\left(\int_{V'} {\mbox{$\Bvec'$}}^2/\mu_0 dV +
              \int_V \rho{\mbox{$\uvec'$}}^2 dV \right) = \\
&- \int_V\big( (\curl \Bvec')^2/(\mu_0^2\sigma) + \rho\nu (\curl \uvec')^2 \big) dV \\
&+ \int_V\! \curl\Bvec'\!\cdot\!(\uvec_0\times\Bvec') dV /\mu_0 - \rho\int_V \!\curl\uvec'\!\cdot\!(\uvec_0\times\uvec') dV
\end{aligned}\label{energy}
\end{equation}
with $V'$ being the infinite space minus any volume with infinite conductivity and
$V$ the volume of the container. 
Of course, 
solutions with growing total energy are impossible, as long as $\uvec_0=\zervec$.
More generally, even if we would admit a rigid body motion for $\uvec_0$, growing solutions 
do not exist.

The situation changes qualitatively, if $\Bvec_0$ is no longer curl--free:
The additional term $- \beta\curl(\curl\Bvec_0\times\Bvec')$ occurring in the
linearized induction equation results in the additional energy term
\begin{equation}
 -\beta\int_V \curl\Bvec'\cdot(\curl\Bvec_0\times\Bvec') dV/\mu_0\;,
 \label{zuterm}
\end{equation}
which quite analogously to the term $\int_V \curl\Bvec'\cdot(\uvec_0\times\Bvec') dV /\mu_0$
is potentially capable of delivering energy.
Hence, the argument concerning the necessity of shear for the occurrence of an instability in the model of RS in fact
supports our findings in \cite{RG02,GR02}, when the term `shear' is no longer used to refer to
macroscopic motions only, but is extended to the microscopic motions of the carriers creating
the current $\curl \Bvec_0/\mu_0$. If the latter should be capable of replacing the shear velocity 
$\uvec_0$, it must not be interpretable as a rigid body motion. 
Therefore, a background field exhibiting
a sufficiently curved profile, is a necessary condition for the
occurrence of the instability we reported on, 
as we stressed in all our papers on this issue.
(A suitable profile for a plane slab $-1\le z\le 1$
with its normal in $z$--direction is, for instance, $\Bvec_0 = \ithat{B}_0(1-z^2) \exvec$
as used in \cite{RG02}.)

\begin{figure}
\hspace*{-.8cm}\epsfig{file=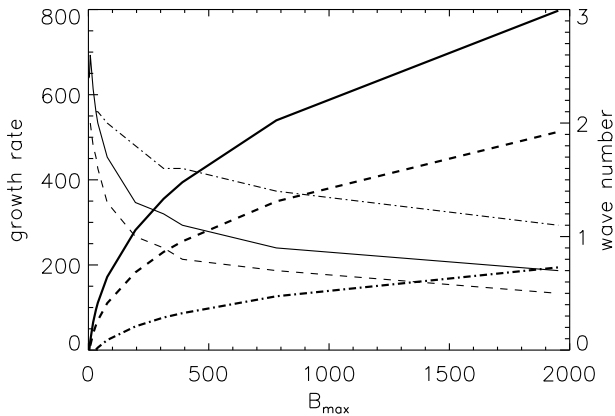,
        width=\linewidth}    
\caption{\label{growrate}
Growth rates (thick) and wave numbers (thin) of the most unstable axisymmetric
magnetic field modes
(kinematic case: $\uvec'=\zervec$) as a function of the background field strength. Length, time and magnetic
field are normalized by $R_\tout$, $R_\tout^2/\eta$ and $\eta/\beta$, respectively.
$B_\tmax$ represents 
the maximum of the background field profile. $ R_\tin/R_\tout = \hat{\eta}
=0.5$. Boundary conditions as defined in
RS; vacuum -- vacuum: solid, inner perfect conductor -- outer vacuum: dashed, perfect
conductor -- perfect conductor: dot--dashed. 
The eigenmodes are non--oscillatory in the first two cases, but oscillatory in
the third.
Interestingly, for the vacuum -- vacuum boundary condition the instability emerges
roughly at $B_\tmax \gtrsim 3$ as in the plane model. 
}  
\end{figure}

As the energy term \eqref{zuterm} contains only magnetic fields, the possibility exists,
that even in the absence of any macroscopic motions ($\uvec'=\uvec_0=\zervec$),
say in a crystallized neutron star crust, nevertheless an instability may occur.
For plane geometry we demonstrated that this possibility is real both for a uniform \cite{RG02} 
and stratified slab \cite{RKG04}. In the cylindrical geometry considered in RS the instability
occurs as well. Figure \ref{growrate} shows normalized growth rates and wave numbers of the
most rapidly growing axisymmetric modes vs. the normalized strength of the
background field. Its 
profile was specified as $\Bvec_0(R) \propto (R-R_\tin)^2 (R-R_\tout)^2 \ezvec$.

The second additional energy term due to an non--potential 
background field, 
\begin{equation}
\int_V \uvec'\cdot(\curl\Bvec_0\times\Bvec')\,dV/\mu_0\,, \label{zweiterm}
\end{equation}
is capable of delivering or consuming
energy, too.  Corresponding instabilities exist and are (for $\uvec_0 = \zervec$)
usually referred to as 
unstable Alfv\'en modes (see, e.g., \cite{C88}). Their nature is obviously MHD,
as \eqref{zweiterm}
vanishes for $\uvec'\!=\zervec$ and/or $\Bvec'\!=\zervec$ .

\section{}
\label{qual} 
Another remark, connected with the above seems to be appropriate. In the Results and
Discussion sections
of RS the impression is given, that the reported instability is primarily
one of the flow. In our opinion, there are good reasons, and moreover even evidences provided
by RS itself to interpret this instability as a primarily
magnetic (and not MHD) one for conditions, in which the flow would be stable otherwise
(that is, for `positive shear' and for `negative shear' with subcritical Reynolds numbers, i.e., for the part of the $\Ry$--$\Ha$ plane beneath the dashed line in Fig. 6
of RS) .

Considering the induction equation including differential rotation and Hall effect with the
velocity perturbations suppressed
(i.e., the kinematic case), one has formally the same equation as that which describes
mean--field dynamos due to
differential rotation and the so--called $\omvec\times\jvec$--effect, see \cite{R69,MP82}.
From these calculations and from qualitative considerations, too, it follows, that 
the sign relation between Hartmann number $\Ha$ and $d\Omega/dR$ reported in RS,
(see, e.g., Sect. III) is just the one necessary for dynamo action,
that is, a {\em magnetic} instability. (Note, that Cowling's theorem does not apply.)
In Sect. III B of RS
a marginal curve in the $\Ry-\Ha$ plane is given for the kinematic case $\uvec'=\zervec$. In that part of the
plane where no hydrodynamic or magneto--rotational instability (MRI) exists, it practically coincides with the
marginal curve of the full system's instability.
Thus, one may suppose, that the velocity perturbations are simply ``enslaved" by the magnetic ones
in cases, in which no instabilities occur
without Hall effect. Since an enslaved $\uvec'$ gives rise to additional dissipation, 
the full system should exhibit smaller growth rates compared with those of the kinematic case.
A hint on this is provided by Fig.~6 of RS showing that for $1\lesssim\Ha\lesssim7$
in the full system a slightly
stronger differential rotation is needed for marginal stability than in the kinematic case.
To judge the nature of the instability the signs of those integrals in \eqref{energy} resulting from the
potentially energy--delivering terms with the calculated eigensolutions inserted
could be inspected.

 When assuming the primarily magnetic character of the instability, its suppression with growing
(absolute value of the) Hartmann number
(cf. Figs. 2--4, 7 and 8 of RS) can be explained by the competition of two 
counteracting effects: On the one hand, growing $|\Ha|$ means growing dominance of the
energy--delivering advection term $\curl\,(\uvec_0\times \Bvec')$ in \eqref{indgl} (by virtue of the
admittedly energetically neutral, but `catalyzing' 
Hall--term) over the dissipation term. But on the other hand, it means also
growing efficiency of the Lorentz force in \eqref{navstok} which causes growing dissipation due to
the enslaved velocity perturbations which drain their energy by virtue of the second advection term 
$\curl\,(\uvec'\times \Bvec_0)$ in \eqref{indgl} out of the magnetic perturbations. 
Hence, the occurrence of a minimum with respect to $\Ha$ is quite natural .   
\section{}
Within the discussion of Fig.~5 of RS (Sect. III A),
it is falsely stated, that the existence of the current--free
marginal solution $B'_R\!=B'_z\!=0\,, \; B'_\phi \propto R^{-1}\,,\; \uvec'\!=\!\zervec$
requires {\em both} boundary conditions to be those
of the perfect conductor. In fact, there is no reason why such a current--free (or vacuum) 
solution could not continue from the inner boundary $R=R_\tin$ on to infinity what means
nothing more than satisfying
the corresponding vacuum condition at the outer rim $R=R_\tout$. 
The necessary and sufficient condition for the existence of this 
vacuum solution everywhere outside the surface $R=R_\tin$ is the existence of a net current
in $z$--direction enclosed
by this surface. Because an outer electromotive force is missing, 
a perfect conductor in the interior of the inner cylinder is needed.
Then, e.g., an arbitrary surface
current can flow without losses and therefore endlessly.

However, the dashed and the dot--dashed curves in Fig.~5 which correspond to the perfectly conducting inner
cylinder are anyway incomprehensible as they both should coincide with the curve $k=0$:
This Figure is intended to show the wavenumber belonging to the 
marginal eigensolution with the minimum Reynolds number for 
a given Hartmann number. In case the inner boundary condition is "perfect
conductor", always the vacuum solution exists, 
which is marginal and allows also $\Re=0$ (or, equivalently, $\uvec_0=0$),
i.e., is associated with the
minimum possible $\Re$. This solution is characterized by $k=0$ and 
the mentioned curves should show that, except, there were other marginal solutions with
$\Re=0$, but $k\ne0$. But such solutions do not exist, because for $\Re=0$ and any $\Ha$ there is
no (potentially) energy--delivering term in the induction equation (see \eqref{energy}) and the only
possibility for a marginal (i.e., non-decaying) solution is the vacuum one.
Thus it appears, that the vacuum solution was for unknown reasons excluded from
the analysis which led to Fig. 5 and can therefore not be referred to to explain it.

However again, even then Fig. 5 is in disagreement with Fig. 2 of RS. The dashed line in the
former should correspond to the solid line in the latter figure which shows no special behavior
at $\Ha=-2$ where $k$ becomes zero in Fig. 5. 
 
\section{}
Considerations of the effect of turbulence on the Hall coefficient do exist
\cite{H68,H71} (see the last paragraph of RS). Perhaps it is appropriate to mention
here a recent revival of
mean--field Hall--electrodynamics in \cite{MGM02} although there only the
effect of the Hall--drift onto the $\alpha$--coefficient is considered.

\bibliographystyle{prsty}

\input sharuecomm2.bbl
\end{document}

%% file: mymath.tex
\newcommand{\deriv}[3]{\frac{#3\hspace*{-.06em} {#1}}{#3\hspace*{.06em} {#2}}}

\newcommand{\order}[2]{\deriv{@,#1}{#2}{d}}
\newcommand{\parder}[2]{\deriv{#1}{#2}{\partial}}

\newcommand{\rezip}[1]{\frac{1}{#1}}

\newcommand{\curl}{\operatorname{curl}}

\newcommand{\lapl}{\operatorname{\Delta}}

\newcommand{\itover}[2]{\,\hspace{.3mm}#1{\!\hspace{-.3mm}#2}}

\newcommand{\ithat}[1]{\itover{\hat}{#1}}

\newcommand{\ovec}[1]{{\mbox{\boldmath $#1$}}}
\newcommand{\exvec}{\ovec{e}_x}

\newcommand{\ezvec}{\ovec{e}_z}
\newcommand{\zervec}{\ovec{0}}
\newcommand{\tmax}{{\text{max}}}

%% file: mymhd.tex
\newcommand{\Bvec}{\ovec{B}}

\newcommand{\jvec}{\ovec{j}}
\newcommand{\uvec}{\ovec{u}}

\newcommand{\omvec}{{\ovec{\omega}}}

\newcommand{\Ry}{\operatorname{\text{Re}}}

\newcommand{\Ha}{\operatorname{\text{Ha}}}